\documentclass[10pt]{article}
\usepackage{amssymb,amsmath,bm,enumerate}
\usepackage{float}
\usepackage{graphicx}
\usepackage[utf8]{inputenc}
\usepackage{url}
\newcommand{\lqqd}{\hfill $\square$\\}

\newtheorem{example}{Example}

\begin{document}
\thispagestyle{empty}
\title{The importance of the change-of-variables integration technique in deriving the PDFs of important probability distributions}

\author{Jorge Pedraza Arpasi
\thanks{Jorge Pedraza Arpasi:
Centro Tecnológico de Alegrete RS, Universidade Federal do Pampa - UNIPAMPA, Brasil,
E-mail: jorgearpasi@unipampa.edu.br.}}
\maketitle
\begin{abstract}
The purpose of this article is to explicitly describe how the integration technique known as the change of variables is used in graduate-level Probability and Statistics courses. Throughout the article, we employ this integration technique seven times across various examples and results. We demonstrate that four applications of this technique are required to derive the probability density function (PDF) of the Student's t-distribution.
\end{abstract}

\textbf{Keywords.- }  Integration, Change of Variables, Gamma function, Beta function, Chi-squared distribution, Student's distribution.

\section{Introduction}

The technique known as change of variables—or substitution of variables—for calculating an integral is taught in calculus courses within engineering programs at universities worldwide \cite{anton,stewart}. At the introductory level—involving integration over intervals of real numbers—most calculus textbooks refer to this technique as integration by substitution; at this initial stage, no special attention is required regarding the absolute value of the derivative of the substituted variable with respect to the new variable.
Subsequently, at more advanced levels—where integration takes place over regions in the $\mathbb{R}^2$ plane or solids in $\mathbb{R}^3$ space—it is taught that the absolute value of the derivative matrix of the coordinate transformation must be calculated; this quantity has a specific name: the Jacobian.

Meanwhile, in introductory statistics and probability courses, it is taught that an event associated with a continuous random variable is essentially an interval of real numbers. Consequently, the probability of such an event can be calculated using an integral where the integrand is a function known as the probability density function (PDF) \cite{montgomery}. At a higher level—involving multiple joint continuous random variables—events correspond to planar regions (for two variables) or solids in three-dimensional space (for three variables). The probability of an event is then the integral of the PDF over that event \cite{leon-garcia}. Each PDF defines a probability distribution, and in the most common cases, each probability distribution has its own corresponding PDF. Theoretically, this implies the existence of probability distributions that lack a PDF or involve complex combinations of PDFs and probability mass functions (PMFs). When a distribution is defined based on a PDF, both the distribution and the PDF share the same name.

Among the most well-known and important distributions are the Gaussian (or Normal) distribution, the Gamma distribution, the Beta distribution, the Chi-squared distribution, and the Student's t-distribution. The Gaussian, Gamma, and Beta distributions are defined based on the Gaussian, Gamma, and Beta functions, respectively. In contrast, the Chi-squared and Student's t-distributions are defined in terms of other distributions. These latter two probability distributions arose to meet practical needs in inferential statistics. Thus, the contribution of this article is to demonstrate the great utility of the change-of-variables technique in deriving the probability density functions (PDFs) for the Chi-squared and Student's t-distributions. To this end, the initial section formally presents the integration-by-change-of-variables technique for the general n-dimensional case, followed by a discussion of the cases where $n=1, 2,$ and $3$. For the one-dimensional case, we explain why it is unnecessary to account for the absolute value of the derivative of the transformation used for the variable substitution.
The two-dimensional case is particularly interesting as it allows for an understanding of the dynamics behind the change-of-variables transformation. Here, one can visualize how straight lines are transformed into curves and quadrilaterals into non-polygonal regions. For instance, with a minimal amount of imagination, one can visualize how the polar transformation—the most well-known planar transformation—converts rectangles into regions resembling segments of circular rings. In this article, we also describe and employ two other planar transformations. One of these, which transforms rectangles into trapezoids, proves useful in demonstrating the formula relating the Beta and Gamma functions. The other planar transformation, which maps rectangles onto quasi-parabolic regions, is fundamental to deriving the PDF of the Student's t-distribution.

Finally, we need to say in this introduction that this article does not offer an original contribution, as all the ideas, discussions, and proofs regarding the technique of integration by change of variables—as well as those concerning probability distributions and probability densities—already appear, albeit in a scattered manner, in various textbooks on calculus, statistics, and probability
\cite{anton}. The purpose of this article is to present and demonstrate, in a single place, the significance of the variable substitution integration technique in the study of important probability distributions. In our view, this serves as a valuable motivational element for the study of calculus in engineering degree programs and related fields.

\section{Integration by the Change of Variables Formula}
Let $\bm{x}=(x_1,x_2,\dots,x_n)\in \mathbb{R}^n$ be a vector variable and $g(\bm{x})$ an integrable function. If there is a bijective transformation $T:R\subset \mathbb{R}^n \rightarrow \mathbb{R}^n$ with $T_i(\bm{u})=x_i$, $\bm{u}=(u_1,u_2,\dots,u_n)$, and Jacobian
\[
JT(\bm{u})=JT(u_1,u_2,\dots,u_n)=\frac{\partial(x_1,x_2,\dots,x_n)}{\partial(u_1,u_2,\dots,u_n)}=
\left| \begin{matrix}
           \frac{\partial x_1}{\partial u_1}&\frac{\partial x_2}{\partial u_1}&\dots&\frac{\partial x_n}{\partial u_1}\\
           \frac{\partial x_1}{\partial u_2}&\frac{\partial x_2}{\partial u_2}&\dots&\frac{\partial x_n}{\partial u_2}\\
           \vdots&\vdots&\vdots&\vdots\\
           \frac{\partial x_1}{\partial u_n}&\frac{\partial x_2}{\partial u_n}&\dots&\frac{\partial x_n}{\partial u_n}\\
          \end{matrix} \right|,
\]
then the integral $\int_R f(\bm{x})d\bm{x}$ can be evaluated by the formula
\begin{equation}\label{eq:change}
 \iint\limits_{R} g(\bm{x}) dV_{\bm{x}} = \iint\limits _S g(T(\bm{u})) JT(\bm{u}) dV_{\bm{u}},
\end{equation}
where $S\in \mathbb{R}^n$ is such that $T(S)=R$ or equivalently $S=T^{-1}(R)$. Some authors describe this technique saying that the ``old'' variable $\bm{x}$ was changed by the ``new'' variable $\bm{u}$. Also, some other authors prefer to call this technique as ``integration by substitution''.

\subsection{The unidimensional case}
In the case $n=1$, the integration by change of variable is where the denomination ``integration by substitution'' has wide acceptation. In this case  the formula (\ref{eq:change}) becomes:
\begin{equation}\label{change1}
 \int\limits_I g(x)dx = \int\limits_J g(T(u)) JT(u) du,
\end{equation}
where $x=T(u)$, $JT(u)=\vert T^\prime (u)\vert$, and  $I=T(J)$. Since $T$ is bijective then $T$ must be monotone which means $T^\prime(u)\neq 0$ for all $u\in J$.\\
Suppose $I=[a,b]$. If $T^\prime(u) >0$ then $(T^{-1})^\prime(x)>0$ and
$T^{-1}(a) < T^{-1}(b)$. Hence $J=T^{-1}(I)=[T^{-1}(a),T^{-1}(b)]$ and
\[\int\limits_I= g(x)dx=\int_a^b g(x) dx= \int\limits_J g(T(u)) \vert T^\prime(u)\vert du=
\int_{T^{-1}(a)}^{T^{-1}(b)} g(T(u)) T^\prime(u)du\]
On the other hand, if $T^\prime(u) <0$ then $T^{-1}(a) > T^{-1}(b)$.
Hence $J=T^{-1}(I)=[T^{-1}(b),T^{-1}(a)]$ and
\begin{multline*}
\int\limits_I= g(x)dx=\int_a^b g(x) dx= \int\limits_J g(T(u)) \vert T^\prime(u) \vert du=
-\int_{T^{-1}(a)}^{T^{-1}(b)} g(T(u)) T^\prime(u)du\\
=\int_{T^{-1}(b)}^{T^{-1}(a)} g(T(u)) T^\prime(u)du\\
\end{multline*}
That is the explanation to the question why in the integration by substitution we do not need to pay attention to the absolute value of $T^\prime(u)$. We had seen that instead of watch the absolute value of the derivative of $T$ it is more practical to care the limits of the interval $J$.

\begin{example}
  Let $g(x)$ be the function $g(x)=\frac{1}{\sqrt{x(1-x)}}$, $0 < x < 1$, then evaluate the integral $\int_0^1 g(x)dx$.
\end{example}
If we define $x=T(u)=\sin^2(u)$, we have that the interval $J=[0,\pi/2]$ is such that $T(J)=[0,1]$ with $T(0)=0$ and $T(\pi/2)=1$ and also
$T^\prime(u)=2\cos(u)\sin(u)>0$ for all $u\in (0,\pi/2)$. Thus
\[\int_0^1 \frac{dx}{\sqrt{x(1-x)}}= \int\limits_0^{\pi/2} \frac{1}{\sqrt{\sin^2(u)\cos^2(u)}}\cdot(2\cos(u)\sin(u))du=
2\int\limits_0^{\pi/2} du= \pi.\]
\lqqd

\subsection{The planar case}
In the case $n=2$ the formula (\ref{eq:change}) becomes
\begin{equation}\label{eq:change2}
 \iint\limits_{R} g(x,y) dA_{xy} = \iint\limits _S g(T(u,v)) JT(u,v) dA_{uv}
\end{equation}

where $(x,y)=T(u,v)$,
the Jacobian $JT$ is
\[
\frac{\partial (x,y)}{\partial (u,v)} = JT(u,v)=\left| \begin{matrix}
           \frac{\partial}{\partial u}(x(u,v))&\frac{\partial}{\partial u}(y(u,v))\\
           \frac{\partial}{\partial v}(x(u,v))&\frac{\partial}{\partial v}(y(u,v)\\
          \end{matrix} \right|,
\]
and $T(S)=R$ or $S=T^{-1}(R)$.

\begin{example}
 The most famous change of variables of the plane is the Polar transformation which changes rectangular system of coordinates in polar coordinates.
 \begin{equation}\label{eq:polar}
 T(\theta,r)=(x(\theta,r)\,,\,y(\theta,r))=(r\cos(\theta)\,,\,r\sin(\theta)).
 \end{equation}

\end{example}

\begin{figure}[h]
 \centering
 \includegraphics[width=12cm]{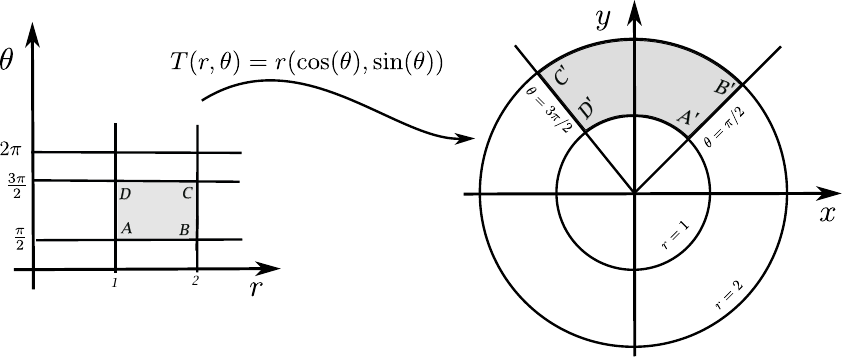}
 \caption{The Polar Transformation}
 \label{fig:polar}
 \end{figure}

 As can be seen in Figure \ref{fig:polar} the rectangular region $ABCD$ inside the $r\theta$-plane  is transformed in the region $A^\prime B^\prime C^\prime D^\prime$ of the $xy$-plane which is a portion of an annulus. The dynamics of this transformation can be described as follows:
 \begin{itemize}
  \item The horizontal line $\theta=\pi/2$ is transformed as $T(r,\pi/2)=r(\cos(\pi/2),\sin(\pi/2))$
  = $\frac{r\sqrt{2}}2(1,1)$ which is the line $y=x$.
  \item The horizontal line $\theta=3\pi/2$ is transformed as $T(r,3\pi/2)=r(\cos(3\pi/2),\sin(3\pi/2))$
  = $\frac{r\sqrt{2}}2(-1,1)$ which is the line $y=-x$.
  \item The vertical line $r=1$ is transformed as $T(1,\theta)= (\cos(\theta),\sin(\theta))$ which is the circle $x^2+y^2=1$.
  \item The vertical line $r=2$ is transformed as $T(1,\theta)= 2(\cos(\theta),\sin(\theta))$ which is the circle $x^2+y^2=4$.
 \end{itemize}

If $R_a$ is the circular region in the $xy$-pĺane given by
\[R_a=\{x^2+y^2 \leq a^2\},\]
then $S=T^{-1}(R_a)$ in the $r\theta$ plane is the rectangle:
\[S=T^{-1}(R_a)=\{0\leq r\leq a\}\times \{0\leq \theta \leq 2\pi\}.\]

The Jacobian $JT(r,\theta)=r$ of polar transformation is:
\[JT(r,\theta)=\frac{\partial(x,y)}{\partial(r,\theta)}= \left\vert \begin{matrix}
                                                                     \cos(\theta)&\sin(\theta)\\
                                                                     -r\sin(\theta)&r\cos(\theta)
                                                                    \end{matrix}\right\vert =r.
\]

Hence
\[\iint\limits_{R_a} g(x,y)dA_{xy}=\iint\limits_S g(r\cos(\theta),r\sin(\theta)) r dA_{r\theta}=
\int\limits_0^{2\pi}\int\limits_0^a g(r\cos(\theta),r\sin(\theta)) rdrd\theta.\]
\lqqd

\subsection{The tridimensional (spatial) case}
In the case $n=3$, the formula (\ref{eq:change}) becomes
\begin{equation}\label{eq:change3}
 \iint\limits_{G} g(x,y,z) dV_{xyz} = \iint\limits _S g(T(u,v,w)) JT(u,v,w) dV_{uvw}
\end{equation}

where $(x,y,z)=T(u,v,w)$,
the Jacobian $JT$ is
\[
\frac{\partial (x,y,z)}{\partial (u,v,w)} = JT(u,v,w)=\left| \begin{matrix}
           \frac{\partial}{\partial u}(x(u,v,w))&\frac{\partial}{\partial u}(y(u,v,w))&\frac{\partial}{\partial u}(z(u,v,w)) \\
           \frac{\partial}{\partial v}(x(u,v,w))&\frac{\partial}{\partial v}(y(u,v,w))&\frac{\partial}{\partial v}(z(u,v,w))\\
           \frac{\partial}{\partial w}(x(u,v,w))&\frac{\partial}{\partial w}(y(u,v,w))&\frac{\partial}{\partial w}(z(u,v,w))\\
          \end{matrix} \right|,
\]
and $T(S)=G$ or $S=T^{-1}(G)$.

\section{The Special Functions Gamma and Beta}
The Gamma function is one of the most important functions in mathematics; it is a classical function. According to Davis \cite{davis}, this function has attracted the attention of the most important mathematicians of all time. Davis states, ``...each generation (of mathematicians) has found something interesting to say about the Gamma function. Perhaps the next generation will do so as well''. This special  function is defined by the integral:

\begin{equation}\label{eq:gamma}
 \Gamma(a)=\int_0^\infty t^{a-1}e^{-t} dt,
\end{equation}
for any real number $a$ in the interval $(0, \infty)$.

For $a+1$, we have
\[\Gamma(a+1)=\int_0^\infty t^{a}e^{-t} dt.\]
Consider now the derivative
\[\frac{\partial}{\partial t}(t^ae^{-t})= at^{a-1}e^{-t}-t^ae^{-t},\]
from where:
\[\int_0^\infty \left[\frac{\partial}{\partial t}(t^ae^{-t})\right]dt= \int_0^\infty at^{a-1}e^{-t}dt -\int_0^\infty t^ae^{-t} dt,\]
then
\[t^ae^{-t}\big\vert_0^\infty = a\Gamma(a)-\Gamma(a+1),\]
which means:
\begin{equation}\label{eq:gamma1}
\Gamma(a+1)=a\Gamma(a).
\end{equation}

For $a=1$ we will have:
\[\Gamma(1)=\int_0^\infty e^{-t}dt=1.\]
Then, by iterative application of (\ref{eq:gamma1}) we obtain:
\[\Gamma(2)=\Gamma(1+1)=1.\Gamma(1)=1.1=1\]
\[\Gamma(3)=\Gamma(2+1)=2.\Gamma(2)=2.1!=2!\]
\[\Gamma(4)=\Gamma(3+1)=3.\Gamma(3)=3.2!=3!\]
\[\Gamma(5)=\Gamma(4+1)=4.\Gamma(4)=4.3!=4!\]

and, in general

\[\Gamma(n+1)=n!\]

On the other hand, fixed $a>0$ and $b>0$, the Beta function  is defined by the integral
\begin{equation}\label{eq:beta}
\mathcal{B}(a,b)=\int_0^1 t^{a-1}(1-t)^{b-1}dt
 \end{equation}

The formula relating the Gamma and Beta function is:

\begin{equation}\label{eq:beta_gamma}
\mathcal{B}(a,b)=\frac{\Gamma(a)\Gamma(b)}{\Gamma(a+b)},
 \end{equation}
which in many calculus and probability textbooks are presented without a proof. We will give a proof
of this formula considering the product  $\Gamma(a)\Gamma(b)$ and showing that it can be computed as a double integral over a region of $xy$-plane. Then we apply the change of variables technique (\ref{eq:change2}) with a specific planar transformation that more or less resembles the polar transformation.\\

\begin{multline*}\Gamma(a)\Gamma(b)=\left(\int_0^\infty x^{a-1}e^{-x}dx\right) \left(\int_0^\infty y^{b-1}e^{-y}dy\right)\\
=\int_0^\infty\int_0^\infty x^{a-1}y^{b-1}e^{-(x+y)}dydx
=\iint\limits_{R} g(x,y) dA_{xy},
 \end{multline*}
where $R$ is the region
\[R=\{(x,y)\;;\; x\geq 0, y\geq 0\},\]
and
\[g(x,y)=x^{a-1}y^{b-1}e^{-(x+y)}.\]

\begin{figure}[h]
 \centering
 \includegraphics[width=12cm]{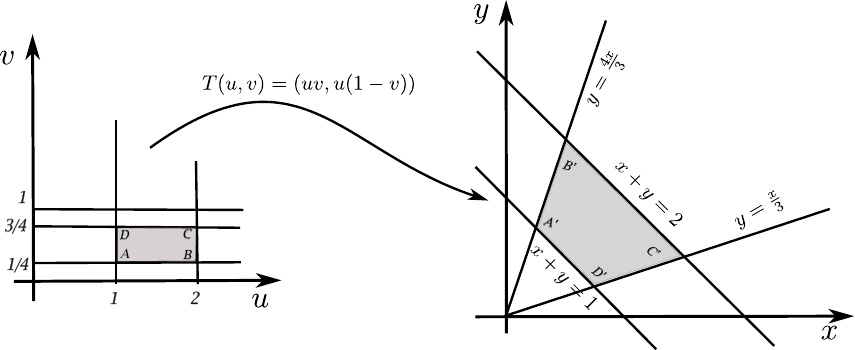}
 \caption{The Trapezoidal Transformation:
 The rectangular region $ABCD$ inside the $uv$-plane  is transformed in the trapezoidal region $A^\prime B^\prime C^\prime D^\prime$ of the $xy$-plane.}
 \label{fig:beta}
 \end{figure}

To apply the  formula (\ref{eq:change2}) consider the convenient mapping $T(u,v)$:
 \begin{equation}\label{eq:trapezoid}
 T(u,v)=(x(u,v),y(u,v))=(uv,u(1-v)),
 \end{equation}

which we call a trapezoidal transformation and whose dynamics can be observed in Figure \ref{fig:beta}.

 We can verify easily that the region $S=T^{-1}(R)$ in the $UV$-plane is the semi-infinite strip:
\[S=\{(u,v)\;;\; 0\leq u\,,\, 0\leq v\leq 1\};\]
while
\begin{multline*}g(T(u,v))=g(uv,u(1-v))
=(uv)^{a-1}(u(1-v))^{b-1}e^{-(uv+u(1-v)}=\\
=u^{a+b-2}e^{-u}v^{a-1}(1-v)^{b-1}.
\end{multline*}
Since the Jacobian of $T$ is $JT(u,v)=u$, then
\[g(T(u,v))\cdot JT(u,v)=u^{a+b-1}e^{-u}v^{a-1}(1-v)^{b-1}.\]
Hence;
\begin{multline*}
 \Gamma(a)\Gamma(b)=\iint\limits_S g(T(u,v))JT(u,v) dA_{uv}
 =\int_0^\infty\int_0^1 u^{a+b-1}e^{-u}v^{a-1}(1-v)^{b-1} dvdu\\
 =\int_0^\infty u^{a+b-1}e^{-u}du \int_0^1 v^{a-1}(1-v)^{b-1} dv
 =\Gamma(a+b)\mathcal{B}(a,b),
\end{multline*}
which effectively proves the famous formula:
\[\mathcal{B}(a,b)=\frac{\Gamma(a)\Gamma(b)}{\Gamma(a+b)}.\]

In the particular case $a=\frac{1}2$ and $b=\frac{1}2$ we have:

\[\mathcal{B}(\tfrac{1}2,\tfrac{1}2)= \frac{\Gamma(\tfrac{1}2)\Gamma(\tfrac{1}2) }{\Gamma(\tfrac{1}2+\tfrac{1}2)}
=\left(\Gamma(\tfrac{1}2)\right)^2.\]

On the other hand, by the Example 1:
\[\pi= \int_0^1 \frac{dt}{\sqrt{t(1-t)}}= \int_0^1 t^{\frac{1}2-1}(1-t)^{\frac{1}2-1}dt = \mathcal{B}(\tfrac{1}2,\tfrac{1}2).\]

Therefore

\[\Gamma(\tfrac{1}{2})=\sqrt{\pi}.\]

With this $\Gamma(\tfrac{3}2)$ = $\Gamma(\tfrac{1}2+1)$ = $\tfrac{1}2\Gamma(\tfrac{1}2)$ = $\tfrac{\sqrt{\pi}}2$ and in general, applying (\ref{eq:gamma1}) for any $n$ odd:

\[\Gamma(\tfrac{n+2}2)=\Gamma(\tfrac{n}2+1)=\tfrac{n}2\Gamma(\tfrac{n}2).\]

\section{The Probability Density Function (PDF) of the Chi-squared distribution}

If $X_1$, $X_2$, ..., $X_n$ are independent normal distributions, where each $X_i$ has PDF
$f(x)=\frac{1}{\sqrt{2\pi}}e^{-\frac{x^2}2}$, $x\in \mathbb{R}$, then the Chi-squared random variable, with $n$ degrees of freedom, is defined as
the sum:
\[U=X_1^2+X_2^2+\dots+X_n^2,\]
and usually it is denoted by the symbol $\chi^2$. The PDF of this random variable is fundamental in the analysis of variance techniques and as we will see in the derivation of the Student's distribution.

First, we will derive the PDFs of $\chi^2$, one-by-one for low values of $n$. That is, $n=1$, $n=2$, and $n=3$.
In the planar and tridimensional cases we will use again the change of variables technique. After that we will see that
there is a more efficient method to find the PDFs of $\chi^2$, for any degree $n$, by doing, again, a simply substitution in the integral that defines the Gamma function.

\vspace{1cm}
\textbf{\underline{The Case $n=1$ .-}} \\

In this case $U=X^2$
\begin{multline*}
F(u)=P(U\leq u ) = P(X^2\leq u) = P(-\sqrt{u}\leq X \leq \sqrt{u})=
\int\limits_{-\sqrt{u}}^{\sqrt{u}} \frac{1}{\sqrt{2\pi}} e^{-\frac{t^2}2} dt \\
=\Phi(\sqrt{u})-\Phi(-\sqrt{u})),
\end{multline*}
where $\Phi(z)=\frac{1}{\sqrt{2\pi}}\int\limits_{-\infty}^z e^{-\frac{t^2}2}dt$.

Hence, the PDF $f(u)=\frac{\partial F(u)}{\partial u}$ is
\begin{equation}\label{eq:chi1}
f(u)=\Phi^\prime(\sqrt{u})\left(\frac{1}{2\sqrt{u}}\right)-\Phi^\prime(-\sqrt{u})\left(\frac{-1}{2\sqrt{u}}\right)
=\frac{e^{-\frac{u}2}}{\sqrt{2\pi u}}.
\end{equation}

\textbf{\underline{The Case $n=2$.-}}\\

In this case $U=X^2+Y^2$ and since $X$ and $Y$ are independent the joint pdf is
\[f_{XY}(x,y)=f_X(x)f_Y(y)= \left(\frac{1}{\sqrt{2\pi}}e^{-\frac{x^2}2}\right) \left(\frac{1}{\sqrt{2\pi}}e^{-\frac{y^2}2}\right)=\frac{1}{2\pi}e^{-\frac{x^2+y^2}2}\]

If $R_u$ is the circular region $R_u=\{x^2+y^2\leq u\}$ of the $xy$-plane,
then
\begin{multline*}
 F(u)=P(U\leq u) =  P(X^2+Y^2\leq u)= \iint\limits_{R_u} f_{XY}(x,y) dA_{xy}
=\frac{1}{2\pi}\iint\limits_{R_u} e^{-\frac{x^2+y^2}2} dA_{xy}.
\end{multline*}
To calculate this integral we will use the polar Transformation (\ref{eq:polar})

We can verify easily that the region $S=T^{-1}(R_u)$ in the $r\theta$-plane is the rectangle:
\[S=\{(r,\theta)\;;\; 0\leq r \leq \sqrt{u}, 0\leq \theta \leq 2\pi\};\]
while
\[f_{XY}(T(r,\theta))=f_{XY}(r\cos \theta,r\sin \theta) = \frac{1}{2\pi}e^{-\frac{r^2}2}\]

Since the Jacobian of $T$ is $JT(r,\theta)=r$, then
\[f_{XY}(T(r,\theta))JT(r,\theta)=\frac{1}{2\pi}re^{-\frac{r^2}2}.\]
Hence applying the change of variables Theorem:
\begin{multline*}
 F(u)=\iint\limits_{R_u} f_{XY}(x,y) dA_{xy} = \iint\limits_S f_{XY}(T(r,\theta))JT(r,\theta)dA_{r\theta}=
 \frac{1}{2\pi}\int_0^{2\pi}\int_0^r re^{-\frac{r^2}2} dr d\theta =
 \int_0^{\sqrt{u}} r e^{-\frac{r^2}2} dr
\end{multline*}

Thus, the PDF $f(u)=\frac{\partial F(u)}{\partial u}$ is
\begin{equation}\label{eq:chi2}
f(u)=\sqrt{u}e^{-\frac{u}2}(\sqrt{u})^\prime=\frac{\sqrt{u}e^{-\frac{u}2}}{2\sqrt{u}}=\frac{1}{2}e^{-\frac{u}2}.
\end{equation}

\textbf{\underline{The Case $n=3$ .-}}\\

In this case $U=X^2+Y^2+Z^2$ and since $X$, $Y$ and $Z$ are independent the joint pdf is
\begin{multline*}
f_{XYZ}(x,y,z)=f_X(x)f_Y(y)f_Y(y)= \left(\frac{1}{\sqrt{2\pi}}e^{-\frac{x^2}2}\right) \left(\frac{1}{\sqrt{2\pi}}e^{-\frac{y^2}2}\right)\left(\frac{1}{\sqrt{2\pi}}e^{-\frac{z^2}2}\right)\\
=\frac{1}{(2\pi)^{3/2}}e^{-\frac{x^2+y^2+z^2}2}
\end{multline*}

If $S_u$ is the spherical solid $S_u=\{x^2+y^2+z^2\leq u\}$ of the $xyz$-space,
then
\begin{multline*}
 F(u)=P(U\leq u) =  P(X^2+Y^2+Z^2\leq u)= \iiint\limits_{S_u} f_{XYZ}(x,y,z) dV_{xyz}\\
=\frac{1}{(2\pi)^{3/2}}\iiint\limits_{S_u} e^{-\frac{x^2+y^2+z^2}2} dV_{xyz}.
\end{multline*}
To calculate this integral we will use the spherical transformation which is a generalization of the polar
transformation
 \begin{equation}\label{eq:spherical}
 T(\rho,\phi,\theta)=(x(\rho,\phi,\theta),y(\rho,\phi,\theta),z(\rho,\phi,\theta)),
 \end{equation}
where
\[x(\rho,\phi,\theta)=\rho\cos(\theta)\sin(\phi),\]
\[y(\rho,\phi,\theta)=\rho\sin(\theta)\sin(\phi),\]
and \[z(\rho,\phi,\theta)=\rho\cos(\phi).\]

We can verify easily that the solid $S=T^{-1}(S_u)$ in the $\rho \phi \theta$-space is the rectangular box:
\[S=\{(\rho,\phi,\theta)\;;\; 0\leq \rho \leq \sqrt{u}\,,\, 0\leq \phi\leq \pi\,,\,0\leq \theta \leq 2\pi\};\]
while
\[f_{XYZ}(T(\rho,\phi,\theta))=f_{XYZ}(\rho\cos(\theta)\sin(\phi),\rho\sin(\theta)\sin(\phi),\rho\cos(\phi)) =
\frac{1}{(2\pi)^{3/2}}e^{-\frac{\rho^2}2}\]

Since the Jacobian of $T$ is $JT(\rho,\phi,\theta)=\rho^2\sin(\phi)$, then
\[f_{XYZ}(T(\rho,\phi,\theta))JT(\rho,\phi,\theta)=\frac{1}{(2\pi)^{3/2}}\rho^2\sin(\phi)e^{-\frac{\rho^2}2}.\]
Hence applying the change of variables Theorem:

\begin{multline*}
F(u)=\iint\limits_{S_u} f_{XYZ}(x,y,z)dV_{xyz}=\iint\limits_{S} f_{XYZ}(T(\rho,\phi,\theta))JT(\rho,\phi,\theta)dV_{\rho \phi \theta}\\
=\frac{1}{(2\pi)^{3/2}}\int_0^{2\pi} \int_0^\pi \int_0^{\sqrt{u}} \rho^2 e^{-\frac{\rho^2}2} \sin\phi d\rho d\phi d\theta
= \frac{1}{\sqrt{2\pi}} \int_0^\pi \sin \phi d\phi \int_0^{\sqrt{u}} \rho^2  e^{-\frac{\rho^2}2} d\rho\\
= \frac{2}{\sqrt{2\pi}} \int_0^{\sqrt{u}} \rho^2  e^{-\frac{\rho^2}2} d\rho
\end{multline*}

Thus, the PDF is $f(u)=\frac{\partial F(u)}{\partial u}$
\begin{equation}\label{eq:chi3}
f(u)=\frac{2}{\sqrt{2\pi}}(u e^{-\frac{u}2})(\sqrt{u})^\prime=
\frac{2}{\sqrt{2\pi}}(u e^{-\frac{u}2})(\sqrt{u})^\prime\left(\frac{1}{2\sqrt{u}}\right)= \frac{\sqrt{u}e^{-\frac{u}2}}{\sqrt{2\pi}}.
\end{equation}

In this way using a generalized spherical transformations which transform spheres $x_1^2+x_2^2+\dots+x_n^2\leq u$ in rectangular boxes,
we could find the density of $\chi^2$ with $n$ degrees of freedom. But there is a  more powerful method, that also uses
a simple change of variable in the integral that defines the Gamma function.

\subsection{Finding the PDF of $\chi^2$ via the Gamma function}
 By doing the change of variable $t=\lambda s$, $\lambda>0$, in (\ref{eq:gamma}), we can see that $\Gamma(\alpha)$ also can be expressed as
 \[\Gamma(\alpha)=  \int_0^\infty \lambda^\alpha s^{\alpha-1} e^{-\lambda s} ds,
 \]

 which allows to define the Gamma distribution of probability with parameters $\alpha$ and $\lambda$. The PDF of this Gamma distribution  is:
 \begin{equation}\label{eq:chipdf}
  f_{\alpha,\lambda}(x)=\frac{\lambda(\lambda x)^{\alpha-1}e^{-\lambda x}}{\Gamma(\alpha)},\quad 0 < x \leq \infty
\end{equation}

If $\alpha=\frac{n}2$ and $\lambda=\frac{1}2$, the PDF (\ref{eq:chipdf}) becomes

\begin{equation}\label{eq:chipdf1}
f_n(x)=\frac{x^{\frac{n}2-1}e^{-\frac{x}2}}{2^{\frac{n}2}\Gamma(\tfrac{n}2)}, \quad 0 < x \leq \infty.
\end{equation}
 For $n=1$, the PDF (\ref{eq:chipdf1}) has the form:
\[\frac{x^{\frac{1}2-1}e^{-\frac{x}2}}{2^{\frac{1}2}\Gamma(\tfrac{1}2)}= \frac{e^{-\frac{x}2}}{\sqrt{2\pi x}},\]
which do coincides with (\ref{eq:chi1}).

 For $n=2$, the PDF (\ref{eq:chipdf1}) has the form:
 \[\frac{x^{\frac{2}2-1}e^{-\frac{x}2}}{2^{\frac{2}2}\Gamma\left(\frac{2}2\right)}= \frac{1}2 e^{-\frac{x}2},\]
which do coincides with (\ref{eq:chi2}).

 For $n=3$, the PDF (\ref{eq:chipdf1}) has the form:
 \[\frac{x^{\frac{3}2-1}e^{-\frac{x}2}}{2^{\frac{3}2}\Gamma\left(\frac{3}2\right)}=\frac{\sqrt{x}e^{-\frac{x}2}}{\sqrt{2\pi}}\]
which do coincides with (\ref{eq:chi3}).

In the Table \ref{tab:chi} we show a list of PDFs of $\chi^2$ for degrees $n=1,2,3,4,5,6$
\begin{table}[H]
\[
 \begin{array}{c|c|c}
  \mbox{degree}\, $n$ &f_n(x)\,\mbox{in}\, (\ref{eq:chipdf1}) & \mbox{Simplified}\,f_n(x)\\ \hline \hline
  1&\frac{x^{\frac{1}2-1}e^{-\frac{x}2}}{2^{\frac{1}2}\Gamma\left(\frac{1}2\right)}&\frac{e^{-\frac{x}2}}{\sqrt{2\pi x}}\\ \hline
  2&\frac{x^{\frac{2}2-1}e^{-\frac{x}2}}{2^{\frac{2}2}\Gamma\left(\frac{2}2\right)}& \frac{1}2 e^{-\frac{x}2}\\ \hline
  3&\frac{x^{\frac{3}2-1}e^{-\frac{x}2}}{2^{\frac{3}2}\Gamma\left(\frac{3}2\right)}&\frac{\sqrt{x}e^{-\frac{x}2}}{\sqrt{2\pi}}\\ \hline
  4&\frac{x^{\frac{4}2-1}e^{-\frac{x}2}}{2^{\frac{4}2}\Gamma\left(\frac{4}2\right)}& \frac{x e^{-\frac{x}2}}{4}\\ \hline
  5&\frac{x^{\frac{5}2-1}e^{-\frac{x}2}}{2^{\frac{5}2}\Gamma\left(\frac{5}2\right)}&\frac{x^{\frac{3}2}e^{-\frac{x}2}}{3\sqrt{2\pi}}\\ \hline
  6&\frac{x^{\frac{6}2-1}e^{-\frac{x}2}}{2^{\frac{6}2}\Gamma\left(\frac{6}2\right)}& \frac{x^2 e^{-\frac{x}2}}{16}\\ \hline
 \end{array}
\]
\caption{PDFs of $\chi^2$ for $n=1,2,3,4,5,6$}
\label{tab:chi}
\end{table}

\section{The PDF of the Student's $t$-Distribution}
Published in 1908 in the journal Biometrika, "The Probable Error of a Mean" \cite{student} is a landmark paper in modern statistics. It was authored by William Sealy Gosset, who famously wrote under the pseudonym "Student" to bypass corporate restrictions. The paper revolutionized data analysis by solving a major flaw in statistical theory: how to draw accurate conclusions about a population when working with small sample sizes. There is an article \cite{zabell} marking the centenary of Gosset's publication that recounts the details, historical context, and theoretical and practical impact of the Student's t-distribution on the field of statistics.

Suppose $f_{XY}(x,y)$ is the PDF of a probability distribution over a region $R\subset \mathbb{R}^2$, that is
\[\iint\limits_{R} f_{XY}(x,y) dA_{xy}=1.\]
If $T(u,v)=(x(u,v),y(u,v))$ is a bijective transformation, with Jacobian $JT(u,v)$, such that there is a region $S\subset \mathbb{R}^2$ such that $S=T^{-1}(R)$,
then by the change o variables Theorem:
\[1=\iint\limits_{R} f_{XY}(x,y) dA_{xy}=\iint\limits_{S} f_{XY}(T(u,v)) JT(u.v) dA_{uv}.\]
We can interpret the above equation saying that
\[f_{UV}(u,v)=f_{XY}(T(u,v)) \cdot JT(u,v), \]
is the PDF over the region $S$ for the new random variables $U$ and $V$.\\
Moreover, if $X$ and $Y$ are independent random variables then
\begin{equation}\label{eq:independent}
 f_{UV}(u,v)=f_{X}(x(u,v)) \cdot f_Y(y(u,v))\cdot JT(u,v)
\end{equation}

\begin{figure}[h]
 \centering
 \includegraphics[width=12cm]{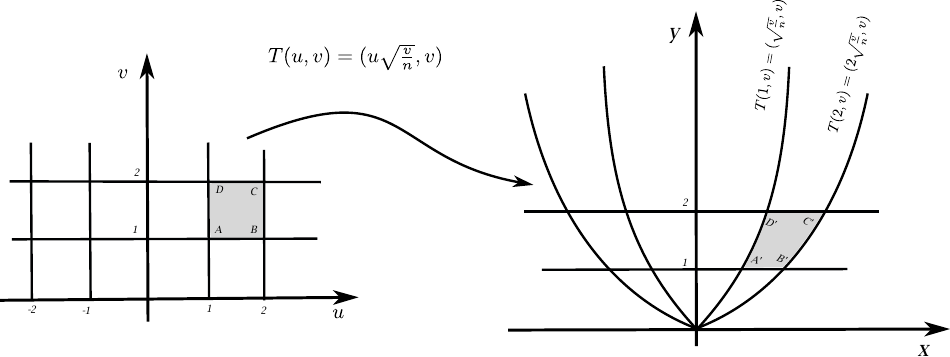}
 \caption{The almost-Parabolic Transformation: The rectangular region $ABCD$ inside the $uv$-plane  is transformed in the  region $A^\prime B^\prime C^\prime D^\prime$ of the $xy$-plane}
 \label{fig:student}
 \end{figure}

To derive the PDF of Student's $t$-distribution, with $n$ degrees of freedom, we will use the equality (\ref{eq:independent}) and the
almost-parabolic transformation:
 \begin{equation}\label{eq:student}
 T(u,v)=(x(u,v),y(u,v))=\left(u\sqrt{\frac{v}n},v\right),
 \end{equation}
whose dynamics can be observed in Figure \ref{fig:student};

The Jacobian of $T(u,v)$ is $JT(u,v)=\sqrt{\frac{v}n}$\\

Now, consider the random variable
\[U=\frac{X}{\sqrt{\frac{Y}n}},\]
where $X$ is the zero-mean, unit-variance Gaussian random variable whose PDF is:
\[f_X(x)=\frac{1}{\sqrt{2\pi}} e^{-\frac{x^2}2},\quad x\in \mathbb{R}\]
and $Y$ is the chi-squared distribution with $n$ degrees of freedom whose PDF was found in (\ref{eq:chipdf1}):

\[f_Y(y)=\frac{y^{\frac{n}2-1}e^{-\frac{y}2}}{2^{\frac{n}2}\Gamma\left(\frac{n}2\right)}, \quad 0\leq x \leq \infty\]
We will show that the random variable $U$ match   the Student's $t$-distribution by finding the closed formula of its PDF.
It can be shown, for instance in \cite{leon-garcia} pp 439, that $X$ and $Y$ are independent. Then, by (\ref{eq:independent}),
$f_{UV}(u,v)=f_{X}(x(u,v)) \cdot f_Y(y(u,v))\cdot JT(u,v)$.\\
Since
\[f_X(x(u,v))=\frac{1}{\sqrt{2\pi}}\exp\left(-\frac{(u\sqrt{v/n})^2}2\right)=\frac{1}{\sqrt{2\pi}}\exp\left(-\frac{u^2v}{2n}\right) \]
and
\[f_Y(y(u,v))=\frac{v^{\frac{n}2-1}e^{-\frac{v}2}}{2^{\frac{n}2}\Gamma\left(\frac{n}2\right)} \]
and
\[JT(u,v)=\sqrt{\frac{v}n},\]
then
\begin{multline*}
f_{UV}(u,v)=f_{X}(x(u,v)) \cdot f_Y(y(u,v))\cdot JT(u,v)= \\
\left(\frac{1}{2\pi}\right) \left(\frac{1}{\sqrt{2^n}\Gamma(n/2)}\right)\left(\frac{1}{\sqrt{n}}\right)
\left(\exp\left(-\frac{u^2v}{2n}\right)\right)\left(v^{\frac{n}2-1} \exp\left(-\frac{v}2\right)\right) \left(v^{\frac{1}2}\right)=\\
=\left(\frac{1}{2^{\frac{n+1}2}\sqrt{n\pi}\Gamma(n/2)}\right)v^{\frac{n-1}2}\exp\left(-\frac{v}2\left(\frac{u^2}n+1\right)\right)
\end{multline*}
The PDF $f_U(u)$, of Student's distribution, is obtained integrating with respect of $v$;
\[f_U(u)=\int\limits_0^\infty f_{UV}(u,v)dv=
\frac{1}{2^{\frac{n+1}2}\sqrt{n\pi}\Gamma(n/2)} \int\limits_0^\infty v^{\frac{n-1}2}\exp\left(-\frac{v}2\left(\frac{u^2}n+1\right)\right)dv\]
Once again, we make a change of variables
\[t=\frac{v}2\left(\frac{u^2}n+1\right),\]
then
\begin{multline*}
 f_U(u)=\left(\frac{1}{2^{\frac{n+1}2}\sqrt{n\pi}\Gamma(n/2)}\right)\left(\frac{2}{\frac{u^2}n+1}\right)^{\frac{n+1}2}
 \int\limits_0^\infty t^{\frac{n-1}2} e^{-t}dt = \\
 \frac{1}{\sqrt{n\pi}\Gamma\left(\frac{n}2\right)\left(\frac{u^2}n+1\right)^{\frac{n+1}2}} \int\limits_0^\infty t^{\frac{n+1}2-1} e^{-t}dt
 =\frac{\Gamma\left(\frac{n+1}2\right)}{\sqrt{n\pi}\Gamma\left(\frac{n}2\right)\left(\frac{u^2}n+1\right)^{\frac{n+1}2}}\\
 =\frac{\Gamma\left(\frac{n+1}2\right)}{\sqrt{n\pi}\Gamma\left(\frac{n}2\right)}
 \left(\frac{1}{\frac{u^2}n+1}\right)^{\frac{n+1}2},
\end{multline*}
which is the PDF of Student's distribution.
In standard statistics and probability literature, the Greek letter $\nu$  is used for degrees of freedom instead of $n$. To align with the notation "Student's t," the probability density function (PDF) uses the variable $t$ instead of $u$. Therefore, the standardized form of the PDF for Student's $t$-distribution is:

\[
f(t)=\frac{\Gamma\left(\frac{\nu+1}2\right)}{\sqrt{\nu\pi}\Gamma\left(\frac{\nu}2\right)}
 \left(\frac{t^2}\nu+1\right)^{-\frac{\nu+1}2},\quad t\in \mathbb{R}.
\]

For $\nu=1$ we have that the Student's distribution becomes the Cauchy distribution:
\[f(t)=\frac{1}{\pi}\left(\frac{1}{t^2+1}\right)\]
and for $\nu=2$ the Student's PDF is
\[f(t)=\frac{1}{2\sqrt{2}}\left(\frac{1}{\frac{t^2}2+1}\right)^{3/2}\]
and for $\nu=3$ the Student's PDF is
\[f(t)=\frac{2}{\pi\sqrt{3}}\left(\frac{1}{\frac{t^2}3+1}\right)^2\]
and so on.

Finally, we prove that
\[
\lim\limits_{\nu\rightarrow \infty} \left\{\frac{\Gamma\left(\frac{\nu+1}2\right)}{\sqrt{\nu\pi}\Gamma\left(\frac{\nu}2\right)}
 \left(\frac{t^2}\nu+1\right)^{-\frac{\nu+1}2}\right\} = \frac{1}{\sqrt{2\pi}}e^{-t^2/2}.
 \]

In effect, writing
\[\left(\frac{t^2}\nu+1\right)^{-\frac{\nu+1}2}=
\left(\left(\frac{t^2}\nu+1\right)^{\frac{\nu}{t^2}}\right)^{\frac{t^2}2(-1+\frac{1}\nu)},\]
we have that
\begin{equation}\label{eq:tz1}
\lim\limits_{\nu \rightarrow \infty}\left\{\left(\frac{t^2}\nu+1\right)^{-\frac{\nu+1}2}\right\}= e^{-t^2/2}.
\end{equation}
On the other hand, by using the Stirling's approximation for large values of $x$;
\[\Gamma(x)= \left(\frac{x}e\right)^x \sqrt{2\pi x},\]
we have
\begin{multline*}
 \frac{\Gamma(\tfrac{\nu+1}2)}{\Gamma(\tfrac{\nu}2)}=\left(\frac{(\nu+1)^{\frac{\nu+1}2}}{\nu^{\frac{\nu}2}}\right)
 \left(\frac{(2e)^{\frac{\nu}2}}{(2e)^{\frac{\nu+1}2}}\right)\left(\frac{\nu+1}\nu\right)^{\frac{1}2}
 =\left(\frac{\nu+1}\nu\right)^{\frac{\nu+1}\nu}(\nu+1)^{\frac{1}2} (2e)^{-\frac{1}2} \left(\frac{\nu+1}\nu\right)^{\frac{1}2}= \\
 =\left(\frac{\nu+1}\nu\right)^{\frac{\nu+1}2}(\nu+1)^{\frac{1}2} (2e)^{-\frac{1}2}.
\end{multline*}

Hence,
\[
 \frac{\Gamma(\tfrac{\nu+1}2)}{\sqrt{\nu\pi}\Gamma(\tfrac{\nu}2)}
 =\frac{1}{\sqrt{\pi}}\left(\frac{\nu+1}\nu\right)^{\frac{\nu+2}2} (2e)^{-\frac{1}2}
 =\frac{1}{\sqrt{2\pi}}\left(\left(\frac{\nu+1}\nu\right)^\nu\right)^{\frac{1}2+\frac{1}{\nu}} e^{-\frac{1}2}
\]
Thus
\begin{multline}\label{eq:tz2}
\lim\limits_{\nu\rightarrow \infty} \left\{\frac{\Gamma(\tfrac{\nu+1}2)}{\sqrt{\nu\pi}\Gamma(\tfrac{\nu}2)}  \right\}
=\lim\limits_{\nu\rightarrow\infty} \left\{ \frac{1}{\sqrt{2\pi}}\left(\left(\frac{\nu+1}\nu\right)^\nu\right)^{\frac{1}2+\frac{1}{\nu}} e^{-\frac{1}2}\right\}=\\
=\frac{1}{\sqrt{2\pi}} e^{\frac{1}2}e^{-\frac{1}2} =\frac{1}{\sqrt{2\pi}}.
\end{multline}
Therefore, from (\ref{eq:tz1}) and (\ref{eq:tz2}) we conclude that, when $\nu$ goes to infinity then the PDF of the Student's distribution goes to the PDF of the zero-mean and variance one Gaussian distribution. That is:

\[
\lim\limits_{\nu\rightarrow \infty} \left\{\frac{\Gamma\left(\frac{\nu+1}2\right)}{\sqrt{\nu\pi}\Gamma\left(\frac{\nu}2\right)}
 \left(\frac{t^2}\nu+1\right)^{-\frac{\nu+1}2}\right\} = \frac{1}{\sqrt{2\pi}}e^{-t^2/2}.
 \]

\bibliography{change}
\bibliographystyle{unsrt}
\end{document}